# Chiral polarizer based on encircling EP

Yanxian Wei[1], Hailong Zhou[1, *], Yuntian Chen[1], Yunhong Ding[2, 3], Jianji Dong[1, #], Xinliang Zhang[1]

[1]*Wuhan National Laboratory for Optoelectronics, School of Optical and Electronic Information, Huazhong University of Science and Technology, Wuhan 430074, China*

[2]*Department of Photonics Engineering, Technical University of Denmark, Lyngby, Denmark*

[3]*SiPhotonIC ApS, Virum Stationsvej 207, 2830 Virum, Denmark*

Corresponding author:[*] hailongzhou@hust.edu.cn; [#]jjdong@mail.hust.edu.cn

**Abstract:** Encircling an exceptional point (EP) in a parity-time (PT) symmetric system has shown great potential for chiral optical devices, such as chiral mode switching for symmetric and anti-symmetric modes. However, the chiral switching for polarization states has never been reported although chiral polarization manipulation has significant applications in imaging, sensing, and communication etc. Here inspired by the anti-PT symmetry, we demonstrate an on-chip chiral polarizer by constructing polarization-coupled anti-PT symmetric system for the first time. The transmission axes of the chiral polarizer are different for forward and backward propagation. A polarization extinction ratio of over 10 dB is achieved for both propagating directions. Moreover, a telecommunication experiment is performed to demonstrate the potential applications in polarization encoding signals. It provides a novel functionality for encircling-an-EP parametric evolution and offer a new approach for on-chip chiral polarization manipulation.

**Introduction:**

Non-Hermitian systems, especially parity-time (PT) symmetric systems, have attracted widespread attention for their fascinating physical properties and broad applications. Fruitful advances have been achieved in the field of PT symmetry[1], such as unidirectional propagation and lasing[2-8], single mode lasing[9-12], sense enhancement[13-18], and optoelectronics oscillator (OEO)[19, 20]. Specially, owing to the non-Hermiticity induced nonadiabatic transitions, chiral mode switching was achieved for symmetric and anti-symmetric modes by encircling the exceptional point (EP) in a PT symmetric system[21-29], exhibiting great potential for chiral devices[7, 27, 30]. However, different from the spatial modes, it is unsuitable to perform the encircling EP evolution in an on-chip PT symmetric system for chiral polarization switching, since the polarization modes are asymmetrical, and the coupling between two orthogonal polarization states needs not only the match of effective indexes of the two polarization modes but also the share of a common parallel polarization component between them. As a result, the chiral polarization switching has not yet been yielded to date. But as a basic property of light, polarization has abundant applications, such as communication[31-33], imaging[34-36] and storage[37]. Thus, chiral polarization switching can arose a lot of fascinating interests on handling polarization information.

As the counterpart of PT symmetric systems, anti-PT symmetric systems were first proposed by Ge and Türeci in 2013[38], conjugated to those observed in PT-symmetric ones[39]. Many novel phenomenon about anti-PT symmetry have been observed such as sensing enhancement[40], coherent perfect absorption lasing[41], heat transfer[42], topological superconductors[43] and chiral dynamics[44]. Different from the PT symmetry, the eigenmodes of anti-PT symmetric systems are asymmetrical when the real parts of the corresponding eigenvalue spited, which exactly matches the polarization mode characteristics. Anti-PT symmetry shows great potential in physics and applications. For practical optical applications, they have harsh requirements of pure imaginary coupling coefficient, which were demonstrated by indirectly dissipative coupling[45], nonlinear coupling[46] and spinning the resonator[47], etc. Benefiting from these breakthroughs, anti-PT symmetry can be heralded as a powerful tool to design optical devices with fascinating properties. Especially, the indirect coupling in anti-PT symmetric systems makes it possible to construct chiral polarization switching.

In this paper, we propose and experimentally demonstrate a chiral polarizer based on anti-PT symmetric system for the first time. Conventional polarizers operate by rejecting undesired polarization, and the transmission axes is the same for bidirectional propagation. Whereas, our chiral polarizer operates by rotating the orthogonal polarization state to the transmission axis[48], and exhibits different transmission axes for forward and backward propagation. We induce a transitional mode between transverse electric (TE) and transverse magnetic (TM) modes and realize the encircling-an-EP parametric evolution in an integrated polarization-coupled anti-PT symmetric system. For arbitrary input polarization states, we achieve a polarization extinction ratio of 10 dB between the transmission of TE and TM modes over a bandwidth from 1550 nm to 1590 nm. Moreover, the application of polarization data formatting is also demonstrated by a communication experiment with a bit rate of 10 Gbit/s on-off keying signals. Our work demonstrates a practicable application for encircling EP in non-Hermitian systems, and provides a novel tool for future polarization manipulating such as data formatting and polarization-multiplexing duplex communication.

## Results
### Principle of chiral polarization switching

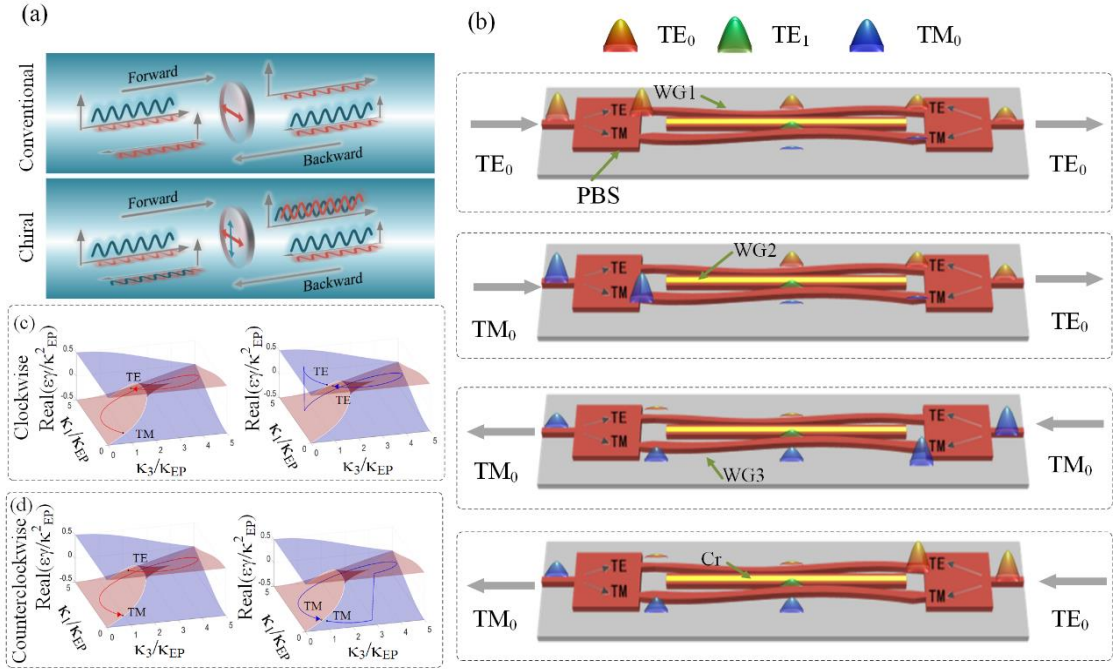

Fig.1. Concept and scheme of the chiral polarizer. (a) The comparison between conventional polarizer and our chiral polarizer. (b) The scheme of our chiral polarizer. The chiral polarizer in construct with three waveguides WG1, WG2 and WG3, where WG1 and WG3 are supposed to be lossless, WG2 has a high loss. (c) Encircling-EP evolution in clockwise. (d) Encircling-EP evolution in counterclockwise.

Fig. 1(a) shows the comparison between conventional polarizer and our chiral polarizer. The chiral polarizer exhibits different transmission axes for forward and backward propagation. In forward propagation, the output polarization state will be rotated to the vertical direction no matter what the input polarization states are. However, in backward propagation, the output polarization state will be rotated to the horizontal direction. This chiral dynamic can be used for on-chip polarization data formatting or polarization multiplexing duplex operation. It should be noted that conventional polarizer operates by rejecting undesired polarization, and the transmission axes is the same for bidirectional propagation. Fig. 1(b) illustrates the chip structure of the chiral polarizer, constructed with a three-waveguide coupled region and two polarization beam splitters (PBSs). The three-waveguide coupled region consists of three waveguides, WG1, WG2 and WG3, where WG1 and WG3 support $TE_0$ and $TM_0$ modes respectively, and WG2 supports the transitional $TE_1$ mode. The coupled region is a dual-port system, in which the $TE_0$ and $TM_0$ mode are processed separately. When light is injected from the left, the output mode is $TE_0$ regardless of $TE_0$ or $TM_0$ injection. Whereas, when light is injected from the right, the output mode is $TM_0$ for both $TE_0$ and $TM_0$ injection. The PBS is used to decompose the input light into $TE_0$ and $TM_0$ mode and combine the output light to handle with arbitrary polarization. The anti-PT symmetry is realized by the indirect dissipate coupling between WG1 and WG3. We use the coupled-mode theory to analyze the system. The model Hamiltonian takes the form:

$$H = \begin{bmatrix} \beta_1 & \kappa_1 & 0 \\ \kappa_1 & \beta_2 + i\gamma & \kappa_3 \\ 0 & \kappa_3 & \beta_3 \end{bmatrix}, \tag{1}$$

where $\gamma$ is the loss of WG2 and it can be tuned by introducing a chromium (Cr) strip on top of $WG_2$, $\beta_1$, $\beta_2$ and $\beta_3$ are the propagation constant of modes in WG1, WG2 and WG3 respectively. $\kappa_1$ is the coupling coefficient between WG1 and WG2, and $\kappa_3$ is that between WG2 and WG3. The eigenvalue is given by $\lambda = \frac{\beta_1 + \beta_3}{2} + i\frac{\kappa_1 \kappa_3}{\gamma} \pm i\sqrt{\frac{\kappa_1^2 \kappa_3^2}{\gamma^2} - \left(\frac{\beta_1 - \beta_3}{2} + i\frac{\kappa_1^2 - \kappa_3^2}{2\gamma}\right)^2}$ with an assumption of $|\gamma| \gg |\lambda - \beta_2|$. The normalized complex-eigenvalue spectra:

$$\varepsilon = \pm i\frac{\kappa_{EP}^2}{\gamma}\sqrt{\frac{\kappa_1^2 \kappa_3^2}{\kappa_{EP}^4} - \left(\text{sign}(\beta_1 - \beta_3) + i\frac{\kappa_1^2 - \kappa_3^2}{2\kappa_{EP}^2}\right)^2} \quad \text{dependent on} \quad \frac{\kappa_1}{\kappa_{EP}} \quad \text{and} \quad \frac{\kappa_3}{\kappa_{EP}} \quad \text{form a Reimann}$$

surface as shown in Figs. 1(c) and 1(d), where the EP point can be specified by $\kappa_1^2 = \kappa_3^2 = \gamma \left|\frac{\beta_1 - \beta_3}{2}\right| = \kappa_{EP}^2$. The encircling-EP evolution can be acquired by suitably changing $\kappa_1$ and $\kappa_3$ along the propagation direction. In the case, chiral polarization switching can be achieved because of the non-Hermiticity induced nonadiabatic transitions[49]. Due to the present of absorption, the self-intersecting Reimann surface is separated into gain surface (red) and loss surface (blue). Each point on the surface represents an eigenvalue of the system. By changing the parameters of the system, eigenvalue can move on the surface. The path can form a ring around the EP if the parameters are well managed, that is, the encircling-EP evolution. The non-adiabatic hopping occurs only when the eigenvalue is moving on the loss surface, which leads to the different final states for encircling EP in clockwise and counterclockwise. Inspired by this chiral dynamic, it is possible to apply encircling-EP evolution to guide the design of chiral polarizer.

Fig. 1(c) shows the process of eigenvalue moving on Reimann surface in clockwise. In this direction, $TE_0$ mode moves on the loss surface. Due to the present of absorption, the evolution of $TE_0$ mode cannot maintain on the loss surface for the entire loop and hops to the gain surface during the evolution. As a result, the $TE_0$ mode returns to itself without mode flipping. However, $TM_0$ mode moves on the gain surface, which indicates that the evolution of $TM_0$ mode can survive to the end without hopping and complete the mode flipping. Fig. 1(d) demonstrate the evolution in counterclockwise. In contrast to the process in clockwise, the $TM_0$ mode moves on the loss surface while the $TE_0$ mode moves on the gain surface when encircling EP in counterclockwise. Figs. 1(c) and 1(d) indicate that, the final polarization state of light for encircling EP in clockwise is $TE_0$ mode and that for encircling EP in counterclockwise is $TM_0$ mode, no matter what original polarization state is.

**Design of the chiral polarizer**

It should be noted that arbitrary polarization state can be decomposed into x and y polarization components with particular amplitude ratio and phase difference. As a result, the input light with arbitrary polarization state is firstly split into $TE_0$ and $TM_0$ modes by a PBS in our device and then starts the encircling-EP evolution. The $TE_0$ and $TM_0$ modes enter the coupled region through WG1 and WG3 respectively after PBS. The length of the coupling region is 110 μm, corresponding to the $L$ in Eq. (2). A longer coupling region length will lead to a higher extinct ratio but a higher loss. A 110 μm coupling length is a good trade-off between extinction ratio and loss. Fig. 2(a) shows the field distributions of $TE_0$ and $TM_0$ modes excited from different directions at 1540 nm. It can be seen clearly that $TE_0$ and $TM_0$ modes are both transmitted to the $TE_0$ mode and output from WG1 when the light is injected from the left ports (i.e., forward propagation). Whereas, they are both transmitted to the $TM_0$ mode and output from WG3 when the light is injected from the right ports (i.e., backward propagation). The corresponding transmission spectra are shown in Fig. 2(b), which indicate that the chiral dynamics occurs over a spectrum range from 1520 nm to 1570 nm. And that a 10 dB extinction ratio is accomplished from 1530 nm to 1560 nm, which can be used for the chiral polarizer.

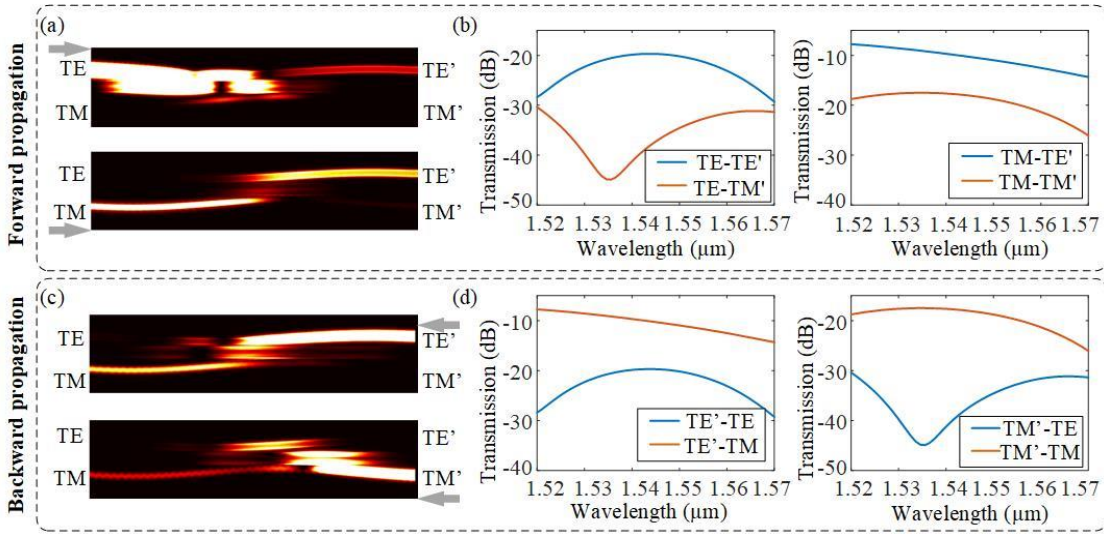

Fig. 2. The simulation results of the chiral polarizer. (a) The simulation power distributions in the waveguides for forward propagation and backward-propagation. (b) The transmission spectra for different input.

**Fabrication and experiment**

The device is fabricated by standard electron-beam (e-beam) lithography based nanofabrication process on the commercial 250 nm SOI platform. The scanning electron microscope (SEM) is used to determine the dimensions of the devices and confirm the obtained geometric shape, as shown in Fig. 3(a). The detailed structures of PBS and coupling region are presented as the insets in Fig. 3(a). The device is characterized by measuring the transmission spectra. The measured transmission spectra for $TE_0$ and $TM_0$ injection are shown in Figs. 3(b-e). TE and TM denote the modes input or output from the left, and TE' and TM' denote the modes input or output from the right. Figs. 3(b) and 3(c) demonstrate the output spectra for forward propagation. It can be seen clearly that the $TE_0$ is the dominant output mode whichever $TE_0$ or $TM_0$ mode is inputted. Whereas, $TM_0$ mode become the dominant output mode for backward propagation, as shown in Figs. 3(d) and 3(e). The oscillations of the experimental spectra are due to the resonances between the facets of the coupling fiber. A 10 dB extinction ratio is achieved from 1550 nm to 1590 nm. The polarizer exhibits a loss of ~25 dB for $TE_0$ inputting from the left and $TM_0$ inputting from the right. This loss is the intrinsic property of encircling EP in passive system. Compared to the simulation results, the experiment spectra are red shifted. This red shift is mainly caused by the fabrication errors.

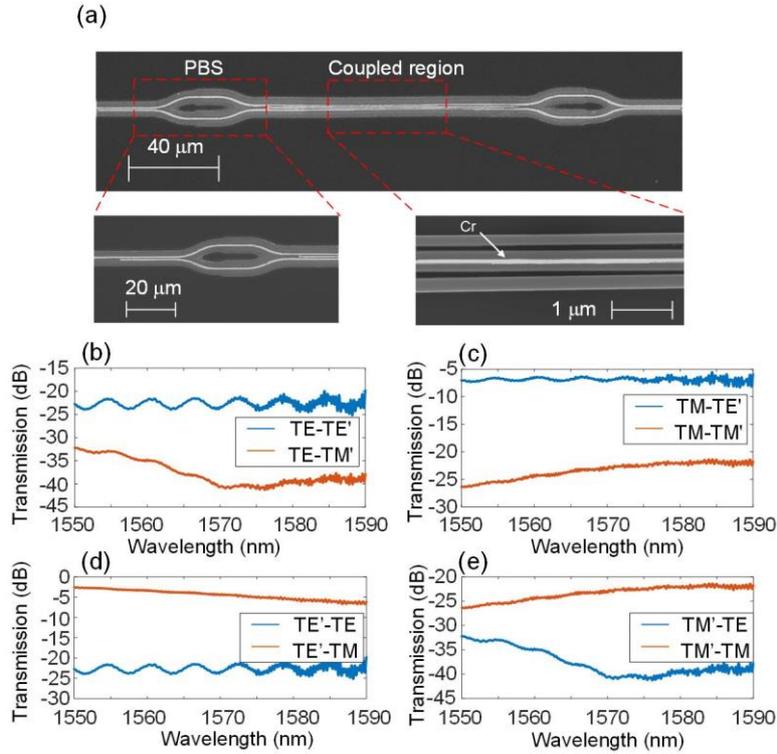

Fig. 3. The SEM images of the chip and the experiment measured spectra. (a) The SEM images of the proposed chiral polarizer. The zoom in images of polarization beam splitter and coupled region are shown in the subgraphs. (b-e) The experimental measured spectra of the device for the (b, c) forward propagation and (d, e) backward-propagation. TE, TM denote the modes input or output from the left, TE' and TM' denotes the modes input or output from the right.

To further verify the practical communication performances of proposed chiral polarizer, a communication experiment is demonstrated to perform a function of formatting the data encoded on polarization channels into a particular polarization state. The experiment setup is demonstrated in Fig. 4(a). The on-off keying (OOK) data are encoded on $TE_0$ and $TM_0$ modes, where $TE_0$ mode indicates '1' and $TM_0$ mode indicates '0'. Without loss of generality, the polarization encoded signals can be decomposed into data loaded on $TE_0$ mode and its invert data loaded on $TM_0$ mode. In the experiment, a 10 GHz random OOK data and its invert data are generated by an arbitrary waveform generator (AWG), and loaded on $TE_0$ and $TM_0$ mode by intensity modulator (IM) respectively. An erbium doped fiber amplifier (EDFA) is used to compensate the power difference between $TE_0$ and $TM_0$ mode induced by the chiral polarizer. Data streams loaded on $TE_0$ and $TM_0$ modes are combined by the PBS and injected into the chip. The output light is decomposed into $TE_0$ and $TM_0$ modes, and received by the oscilloscope (OSC). The input waveforms for forward propagation and backward propagation are shown in Figs. 4(b) and 4(d) respectively. The output $TE_0$ and $TM_0$ polarization components of light are shown in Figs. 4(c) and 4(e). According to the analysis above, the data stream loaded on different polarization states transporting forward will be formatted into $TE_0$ mode. It can be seen clearly in Fig. 4(c) that most signal power is transferred to the $TE_0$ mode, while only a little signal power remains in the $TM_0$ mode. However, the $TM_0$ mode become the dominant mode when the data transports backward. Fig. 4(e) shows that most signal power transferred to $TM_0$ mode rather than $TE_0$ mode, which indicates the function of polarization data formatting with our chiral polarizer. An extinction ratio of 13 dB and 14 dB for forward and backward propagation are achieved respectively.

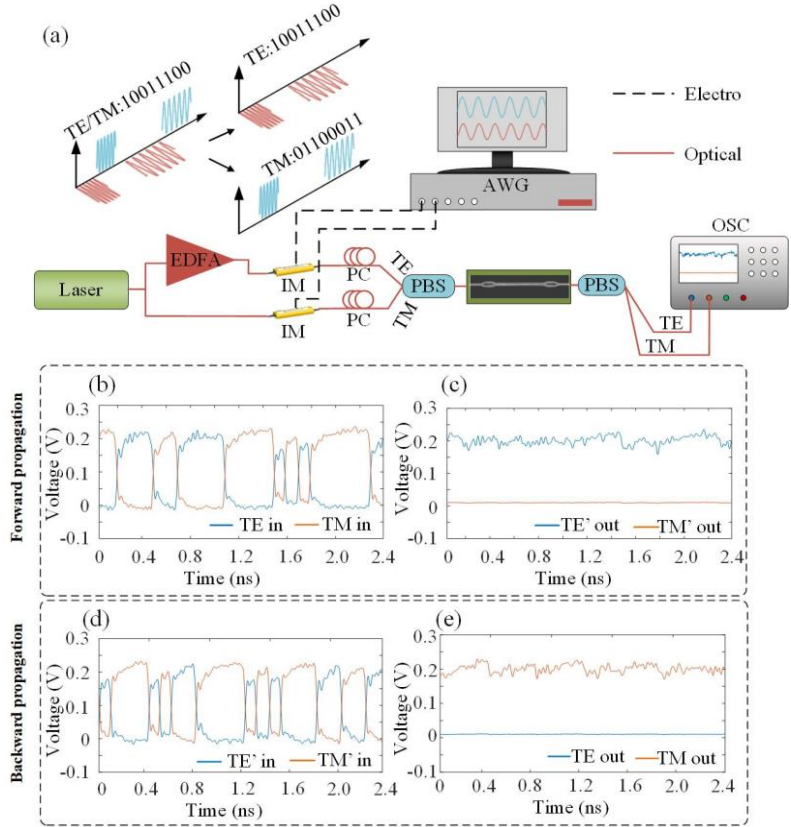

Fig. 4. The experiment setup and results of communication experiment. (a) The experiment setup. (b) The 10 GHz OOK bits stream loaded on TE and TM mode for inputting forward. (c) The power summery of the output TE and TM modes. (d, e) The result of inputting backward. TE, TM denote the modes input or output from the left, TE' and TM' denotes the modes input or output from the right.

**Conclusion:**

In summary, we have proposed a chiral polarizer in an anti-PT symmetric system, which forms different polarized states dependent on light propagation direction. Based on the indirect coupling property of anti-

PT symmetric system, we successfully build up an on-chip polarization-coupled anti-PT symmetric system for the first time. With the encircling-EP evolution, we implement the chiral asymmetry polarization switch, and further the chiral polarizer. The proposed chiral polarizer has successfully applied to polarization data formatting for polarization encoding signals. The extinction ratio between $TE_0$ and $TM_0$ mode for forward and backward propagation is 13 dB and 14 dB respectively. Our work provides a new on-chip polarization manipulation method and demonstrates a practicable application in optics for encircling EP in non-Hermitian systems.

**Methods：**

The chip is fabricated on the commercial 250 nm SOI platform by the e-beam lithography (EBL) based nanofabrication process. First, CSAR 6200 positive e-beam resist is spin-coated on a SOI wafer, and EBL is used to define the etching mask for silicon patterns, which are obtained by Inductively Coupled Plasma (ICP) etching process with a Bosch etching process. Afterwards, a second e-beam lithography is applied to define the chromium (Cr) strip patterns. Finally, electron beam evaporation is used to deposit 100 nm chromium deposition, and Cr strips are obtained by liftoff process.

**Acknowledgements:**
We thank Prof. Chengwei Qiu from National University of Singapore for very fruitful discussions. This work was partially supported by the National Key Research and Development Project of China (2018YFB2201901), the National Natural Science Foundation of China (61805090, 62075075).

**Authors' contributions**

Y.X.W. and H.L.Z. conceived the idea. Y.X.W. performed the numerical simulations and designed the device. Y.X.W. performed the experiments. Y.H.D. fabricated the chip. Y.X.W., Y.T.C., Y.H.D., discussed the results. J.J.D and X.L.Z supervised the study. All authors contributed to the writing of the paper.

**Data availability**

All the data related to this paper are available from the corresponding authors upon request.

**Additional informati**

**Competing financial interests:** The authors declare no competing financial interests.